\documentstyle[12pt]{article}

\begin{document}
\begin{titlepage}

\title{On gravitational radiation and the energy flux of matter}

\author{J. W. Maluf$\,^{*}$ and F. F. Faria\\
Instituto de F\'{\i}sica, \\
Universidade de Bras\'{\i}lia\\
C. P. 04385 \\
70.919-970 Bras\'{\i}lia DF, Brazil\\}
\date{}
\maketitle

\begin{abstract}

A suitable derivative of Einstein's equations in the
framework of the teleparallel equivalent of general relativity
(TEGR) yields
a continuity equation for the gravitational energy-momentum.
In particular, the time derivative of the total gravitational
energy is given by the sum of the total fluxes of gravitational and
matter fields energy. We carry out a detalied analysis of the
continuity equation in the context of Bondi and Vaidya's 
metrics. In the former space-time the flux of gravitational energy is
given by the well known expression in terms of the square of the news
function. It is known that the energy definition in the realm of the
TEGR yields the ADM (Arnowitt-Deser-Misner) energy for appropriate
boundary conditions. Here we show that the same energy definition
also describes the Bondi energy. The analysis of the continuity
equation in Vaidya's space-time shows that the variation of the total
gravitational energy is determined by the energy flux of matter only.

\end{abstract}
\thispagestyle{empty}
\vfill
\par
\bigskip
\noindent (*) e-mail: wadih@fis.unb.br\par
\end{titlepage}
\newpage

\noindent

\section{Introduction}
\bigskip

In spite of its wide acceptance as a theory for the dynamics of the
space-time and of the gravitational field, Einstein's theory of
general relativity, in the metrical formulation, does not allow the
emergence of an unequivocal definition of gravitational energy.
The definition of such quantity is surely important for a
comprehensive understanding of the theory. In the past the major
attempts at defining the gravitational energy were carried out by
means of pseudotensors \cite{Landau,Weinberg,Goldberg}.
Nowadays the concept of quasilocal energy
(i.e., the energy associated to a closed spacelike 2-surface) is being
thoroughly investigated in the context of Hamiltonian formulations
(see, for instance, Ref. \cite{Nester1} and references therein).
Although this approach is promising, there is no ultimate
consensus about it.

Two well established notions of gravitational energy are the total
energy of the space-time, known as the ADM (Arnowitt-Deser-Misner)
energy \cite{ADM}, evaluated on a spacelike slice of a space-time
that is asymptotically flat, and the Bondi energy
\cite{Bondi} that describes the mass of radiating systems in
asymptotically flat space-times. The latter is usually defined at
null infinity. A remarkable feature of the Bondi energy is that it is
related (by means of the field equations) to the loss of the total
mass of a source that radiates gravitational waves. The loss of total
mass is given in terms of minus the square of the news function, and
therefore the mass of the source can only decrease.

A definition for the gravitational energy-momentum has been
investigated in the realm of the teleparallel equivalent of general
relativity (TEGR) \cite{Maluf1,Maluf2} (Ref. \cite{Hehl}
considers the TEGR as a particular case of metric-affine theories of
gravity, and provides a large number of references on the subject).
Attempts to define the
gravitational energy in the context of teleparallel theories of
gravity were first considered by M\o ller \cite{Mol}, who noticed
that an expression for the gravitational energy density could
be constructed out of the torsion tensor. Arguments in favour of the
localizability of the gravitational energy are presented in Refs.
\cite{Maluf1,Maluf2,Maluf3}, where it is argued that it is possible
to identify the gravitational energy density with an appropriate
scalar density, defined in terms of a set of global tetrad
fields, which is a total divergence. Therefore the gravitational
energy density in consideration transforms as a scalar density
under coordinate transformations. We recall that the localizability
of the gravitational energy was already supported by Bondi
\cite{Bondi2}, who analyzed the possible forms of energy transfer
of the gravitational field. He claimed that in relativity a
nonlocalizable form of energy is inadmissible, because all forms of
energy contribute to gravity, and therefore its
location can in principle be found.

However, it is reasonable to expect the {\it localization} of
the gravitational energy to depend on the reference frame
\cite{Maluf3}. The Principle of Equivalence is based on the equality
of inertial and gravitational masses. Alternatively, an accelerated
frame can be locally considered as a rest frame with the addition of a
certain gravitational field. Therefere our perception of the strength
of the gravitational field on nearby bodies depends clearly on our
reference frame, and so does the gravitational energy as measured on
the same frame. Reference frames are described in terms of tetrad
fields \cite{Synge,Aldrovandi}, which are the basic field quantities
of the TEGR. The total energy of a space-time, however, is not
expected to depend on a particular frame (the total gravitational
energy of any asymptotically flat space-time, in the TEGR,
is precisely the ADM
energy \cite{Maluf1}). The dependence of the localization of the
gravitational energy on the frame is a manifestation of the
fact that gravitational and inertial forces are of the same nature.

In this article we address the issue of the variation of the
gravitational energy and the corresponding total flux of energy at
spacelike infinity. The success of the present analysis may qualify
the TEGR as an adequate framework to investigate more intricate
astrophysical models of energy flux and gravitational waves
emission. We conclude that the energy-momentum definition
in consideration here provides a unique framework to analyze both
the ADM and the Bondi energy. We will show that for the Bondi
space-time metric tensor the decrease in time of the total
gravitational energy causes the flux of gravitational energy
only (by means of the emission of gravitational waves), whereas in
Vaidya's space-time \cite{Vaidya,Lind} the loss of gravitational
energy is related to a pure flux of matter energy. The variation in
time of the gravitational energy is determined by the sum of the
gravitational and matter energy fluxes. From the point of view of
Bondi \cite{Bondi2}, these fluxes characterize the intangible and
tangible forms of gravitational energy transfer, respectively
(the latter being determined by the energy-momentum tensor).

In section II of the article we recall the
simple derivation of the continuity equation for the gravitational
energy-momentum. In section III we present the construction of the
simplest set of tetrad fields, defined in terms of the retarded time
$u=t-r$, that describes the Bondi space-time. The Bondi energy, the
loss of total mass and the total flux of gravitational energy at
spacelike infinity are obtained in section IV. In section V we
consider the Vaidya space-time metric tensor and relate the loss of
the total mass of the source to a pure matter energy flux. Finally
in section VI we present our conclusions.

Notation: space-time indices $\mu, \nu, ...$ and SO(3,1)
indices $a, b, ...$ run from 0 to 3. Time and space indices are
indicated according to
$\mu=0,i,\;\;a=(0),(i)$. The tetrad field $e^a\,_\mu$ 
yields the definition of the torsion tensor:  
$T^a\,_{\mu \nu}=\partial_\mu e^a\,_\nu-\partial_\nu e^a\,_\mu$.
The flat, Minkowski space-time  metric is fixed by
$\eta_{ab}=e_{a\mu} e_{b\nu}g^{\mu\nu}= (-+++)$.        

\bigskip

\section{The fluxes of gravitational and matter fields energy}

The Lagrangian density for the graviational field in the TEGR, in
the presence of matter fields, is given by

\begin{eqnarray}
L(e_{a\mu})&=& -k\,e\,({1\over 4}T^{abc}T_{abc}+
{1\over 2} T^{abc}T_{bac} -T^aT_a) -L_m\nonumber \\
&\equiv&-k\,e \Sigma^{abc}T_{abc}-L_m\;,
\label{1}
\end{eqnarray}
where $k=1/(16 \pi)$ and $e=\det(e^a\,_\mu)$. The tensor
$\Sigma^{abc}$ is defined by

\begin{equation}
\Sigma^{abc}={1\over 4} (T^{abc}+T^{bac}-T^{cab})
+{1\over 2}( \eta^{ac}T^b-\eta^{ab}T^c)\;,
\label{2}
\end{equation}
and $T^a=T^b\,_b\,^a$. The quadratic combination
$\Sigma^{abc}T_{abc}$ is proportional to the scalar curvature
$R(e)$, except for a total divergence \cite{Maluf4}. $L_m$
represents the Lagrangian density for matter fields. The field
equations for the tetrad field read

\begin{equation}
e_{a\lambda}e_{b\mu}\partial_\nu(e\Sigma^{b\lambda \nu})-
e(\Sigma^{b \nu}\,_aT_{b\nu \mu}-
{1\over 4}e_{a\mu}T_{bcd}\Sigma^{bcd})
\;=\;{1\over {4k}}eT_{a\mu} \;,
\label{3}
\end{equation}
where

$$
{{\delta L_m}\over {\delta e^{a\mu}}}\equiv eT_{a\mu}\;.
$$
It is possible to prove by explicit calculations that the left hand
side of Eq. (3) is exactly given by ${1\over 2}\,e\,
\lbrack R_{a\mu}(e)-{1\over 2}e_{a\mu}R(e)\rbrack$.

Multiplication of Eq. (3) by the appropriate inverse tetrad fields
yields the latter equation in the form

\begin{equation}
\partial_\nu(-4ke\Sigma^{a\lambda \nu})=
-ke e^{a\mu}(4\Sigma^{b\lambda \nu}T_{b\nu\mu}-
\delta^\lambda_\mu \Sigma^{bdc}T_{bcd})-ee^a\,_\mu T^{\lambda \mu}\;.
\label{4}
\end{equation}
As usual, tetrad fields convert space-time into Lorentz indices and
vice-versa.
By making $\lambda=j$ (i.e., by restricting the space-time index
$\lambda$ to assume only space values) we obtain

\begin{eqnarray}
-\partial_0(-4ke\Sigma^{a0j})-\partial_k(-4ke\Sigma^{akj})=
&-&kee^{a\mu}(4\Sigma^{bcj}T_{bc\mu}-
\delta^j_\mu\Sigma^{bcd}T_{bcd})\nonumber \\
&-&ee^a\,_\mu T^{j\mu}\;.
\label{5}
\end{eqnarray}
Now we take the derivative of the equation above with respect to
$j$. The second term on the left hand side vanishes because of
the antisymmetry property  $\Sigma^{akj}=-\Sigma^{ajk}$. The
resulting equation reads

\begin{eqnarray}
-\partial_0\partial_j(-4ke\Sigma^{a0j})=
&-&k\partial_j\lbrack e e^{a\mu}(4\Sigma^{bcj}T_{bc\mu}-
\delta^j_\mu\Sigma^{bcd}T_{bcd})\rbrack \nonumber \\
&-& \partial_j(ee^a\,_\mu T^{j\mu})\;.
\label{6}
\end{eqnarray}

We recall that in the Hamiltonian formulation of the TEGR
\cite{Maluf5} the momentum canonically conjugated to the tetrad
components $e_{aj}$ is given  by $\Pi^{aj}=-4ke\Sigma^{a0j}$, and
that the gravitational energy-momentum $P^a$ contained within a
volume $V$ of the three-dimensional spacelike hypersurface is
defined by \cite{Maluf1}

\begin{equation}
P^a=-\int_V d^3x\, \partial_j \Pi^{aj}\;.
\label{7}
\end{equation}
If no condition is imposed on the tetrad fields, $P^a$ transforms
as a vector under the global SO(3,1) group. (The Hamiltonian
formulation of the TEGR has also been extensively  discussed by
Blagojevi\'c and Nikoli\'c \cite{BN}.)

By integrating Eq. (6) on a volume $V$ of the
three-dimensional space we arrive at

\begin{equation}
{d \over {dt}}\biggl[ -\int_V d^3x\,\partial_j \Pi^{aj} \biggr]=
-\Phi^a_g -\Phi^a_m\;,
\label{8}
\end{equation}
where

\begin{equation}
\Phi^a_g=k\int_S dS_j\lbrack e e^{a\mu}
(4\Sigma^{bcj}T_{bc\mu}-\delta^j_\mu \Sigma^{bcd}T_{bcd})\rbrack\;,
\label{9}
\end{equation}
is $a$ component of the  gravitational energy-momentum flux, and

\begin{equation}
\Phi^a_m=\int_S dS_j\,(ee^a\,_\mu T^{j\mu})\,,
\label{10}
\end{equation}
is the $a$ component of the matter energy-momentum flux. $S$
represents the spatial boundary of the volume $V$. Therefore the
loss of gravitational energy $P^{(0)}=E$ is governed by the equation

\begin{equation}
{{dE}\over {dt}}=-\Phi^{(0)}_g-\Phi^{(0)}_m\;.
\label{11}
\end{equation}

Equation (8) has been derived previously in the Lagrangian
formulation in Ref. \cite{Maluf3}, without the addition of matter
fields. It does not depend on details of any Hamiltonian formulation.
The time variable $t$ may also stand for the retarded time $u$. We
note that in the analysis of the Hamiltonian formulation of the
gravitational field in null surfaces, in the framework of the TEGR
\cite{Maluf6}, the energy definition given by Eq. (7) was already
anticipated. However the
investigation of the gravitational energy for the Bondi space-time
metric tensor was first carried out by imposing the time gauge
condition \cite{Maluf7} (use was {\it not} made of the retarded
time), a framework that is not quite suitable for such
analysis. In the following sections we will show that $\Phi^{(0)}_g$
plays a major role to Bondi's radiating space-time, since it yields
the well known expression for the gravitational energy loss, whereas
only $\Phi^{(0)}_m$ is relevant to Vaidya's space-time.

\bigskip

\section{The tetrad field for the Bondi space-time}

The tetrad field for the flat space-time, in cartesian coordinates,
is given by $e^a\,_\mu(ct, x,y,z)=\delta^a_\mu$. By making a
coordinate transformation $t \rightarrow u$, where
$u=t-r/c$ is the retarded time and $r=\sqrt{x^2+y^2+z^2}$, we obtain

\begin{equation}
e^a\,_\mu(u,x,y,z)=\pmatrix{1&{x / {rc}}&
{y / {rc}}&{z / {rc}}\cr
0&1&0&0\cr
0&0&1 &0\cr
0&0&0&1\cr}\;,
\label{12}
\end{equation}
where $a$ and $\mu$ label the line and column indices, respectively.
In spherical coordinates we have

\begin{equation}
e^a\,_\mu(u,r,\theta,\phi)=\pmatrix{1&1/c&0&0\cr
0&\sin\theta\cos\phi & r\cos\theta \cos\phi&-r\sin\theta\sin\phi\cr
0&\sin\theta\sin\phi & r\cos\theta \sin\phi& r\sin\theta\cos\phi\cr
0&\cos\theta& -r\sin\theta& 0\cr}\;.
\label{13}
\end{equation}
In Eqs. (12) and (13) only $c$ stands for the velocity of light.

The set of tetrad fields given by Eq. (12) defines a holonomic
transformation $dq^a=e^a\,_\mu dx^\mu$ between the coordinates
$q^{(0)}=u+r/c$, $q^{(1)}=x$, $q^{(2)}=y$ and $q^{(3)}=z$ that
describe a reference space-time \cite{Maluf1}, and the space-time
coordinates $(u,x,y,z)$. In the form given by Eq. (12) the tetrad
fields violate the time gauge condition because $e_{(0)i} \ne 0$,
and therefore $e_{(i)}\,^0 \ne 0$. A nontrivial manifestation of
the gravitational field takes place when the transformation
$dq^a=e^a\,_\mu dx^\mu$ becomes anholonomic \cite{Maluf1}.

It is known that the use of the retarded time $u$ clearly simplifies
the description of the Bondi space-time, since in this case the null
surface condition $g^{00}=0$ can be easily achieved. However it is
not possible to impose at the same time the condition $g^{00}=0$ and
the time gauge condition.

In the flat space-time the violation of the time gauge condition
may be associated to a boost of the physical space-time with
coordinates $x^\mu$ with respect to the reference space-time with
coordinates $q^a$ \cite{Maluf1}. The boost singles out a particular
direction in space. If the time gauge condition is to be violated
in the description of a radiating gravitational field,
then it should not privilege any direction in space, and ideally the
violation should be isotropic. This feature is manifest in the flat
space-time tetrad field given by Eq. (12), and it will
characterize the simplest set of tetrad fields that describes the
Bondi space-time metric tensor, as we will see.

The Bondi metric tensor is given by

\begin{eqnarray}
ds^2=&-&\biggl({V\over r}e^{2\beta}-U^2r^2e^{2\gamma}\biggr)du^2-
2e^{2\beta}\,du\,dr \nonumber \\
&-&2Ur^2e^{2\gamma}\,du\,d\theta+
r^2\biggl(e^{2\gamma}d\theta^2+
e^{-2\gamma}\sin^2\theta\,d\phi^2\biggr)\;,
\label{14}
\end{eqnarray}
where $(r,\theta,\phi)$ are the usual spherical coordinates and
$u$ is the retarded time. The functions in the equation above have
the following asymptotic behaviour

\begin{eqnarray}
\beta &=&-{c^2 \over {4r^2}}+ \cdot \cdot \cdot \nonumber \\
\gamma &=& {c\over r}+ \cdot \cdot \cdot \nonumber \\
{V \over r}&=& 1-{{2M}\over r} \nonumber \\
{}& & -{1 \over r^2} \biggl[
{{\partial d}\over {\partial \theta}}+ d\,\cos\theta-
\biggl({{\partial c}\over{\partial \theta}}\biggr)^2-
4c\biggl({{\partial c}\over {\partial \theta}}\biggr)
\cot\theta \nonumber \\
& &-{1\over 2}c^2\biggl(1+8\cot^2\theta\biggr)\biggr]+
\cdot\cdot\cdot \nonumber \\
U&=& -{1\over r^2} \biggl( {{\partial c}\over{\partial \theta}}
+2c\cot\theta\biggr) \nonumber \\
& & +{1\over r^3}\biggl( 2d +3c {{\partial c}\over{\partial \theta}}
\cot\theta+ 4c^2\cot\theta \biggr)+ \cdot \cdot \cdot \;,
\label{15}
\end{eqnarray}
where $M=M(u,\theta)$ and $d=d(u,\theta)$ are the mass aspect and
the dipole aspect, respectively, and from the function $c(u,\theta)$
we define the news function
$\dot c =(\partial c(u,\theta))/\partial u$. The inverse metric
tensor is given by

\begin{equation}
g^{\mu\nu}(u,r,\theta,\phi)=\pmatrix{0&-e^{-2\beta}&0&0\cr
-e^{-2\beta}& e^{-2\beta}{V\over r}& -e^{-2\beta} U&0\cr
0& -e^{-2\beta} U& {{e^{-2\gamma}} \over r^2}& 0\cr
0&0&0&{{e^{2\gamma}}\over {r^2 \sin^2\theta}} \cr }\;,
\label{16}
\end{equation}
and the determinant of $e^a\,_\mu$ reads $e=\sqrt{-g}
=e^{2\beta}r^2\sin^2\theta$.

The simplest set of tetrad fields $e_{a\mu}(u,r,\theta,\phi)$
that leads to Eq. (14) is
precisely the one for which the violation of the time gauge
condition is isotropic. It reads

\begin{equation}
e_{a\mu}=\pmatrix{
-A&-B&0&0\cr
C\cos\theta\,\cos\phi&B\sin\theta\,\cos\phi
&re{^\gamma}\cos\theta\,\cos\phi&-re^{-\gamma}\sin\theta\,\sin\phi\cr
C\cos\theta\,\sin\phi&B\sin\theta\,\sin\phi
&re^{\gamma}\cos\theta\,\sin\phi& re^{-\gamma}\sin\theta\,\cos\phi\cr
-C\sin\theta& B\cos\theta&-re^{\gamma} \sin\theta&0\cr}\;.
\label{17}
\end{equation}
The functions $A$, $B$ and $C$ are defined by

\begin{eqnarray}
A&=& e^\beta\biggl({V\over r}\biggr)^{1\over 2}\;, \nonumber \\
B&=& e^\beta\biggl({V\over r}\biggr)^{-{1\over 2}}\;, \nonumber \\
C&=&-Ure^\gamma\;.
\label{18}
\end{eqnarray}
It is possible to show by means of a coordinate transformation that

\begin{eqnarray}
e^{(0)}\,_1(u,x,y,z)&=&{x\over r} e^\beta
\biggl({V\over r}\biggr)^{-{1\over 2}} \nonumber \\
e^{(0)}\,_2(u,x,y,z)&=&{y\over r} e^\beta
\biggl({V\over r}\biggr)^{-{1\over 2}} \nonumber \\
e^{(0)}\,_3(u,x,y,z)&=&{z\over r} e^\beta
\biggl({V\over r}\biggr)^{-{1\over 2}}\;.
\label{19}
\end{eqnarray}
Therefore the feature above mentioned regarding the violation of
the time gauge condition is manifest in Eq. (19)
(of course the functions $\beta$ and $V$ depend on $\theta$).
In this respect Eq. (19) shares a similarity with Eq. (12).
Moreover the spatial components of $e_{a\mu}$ in Eq. (17), in
cartesian coordinates, satisfy

\begin{equation}
e_{(i)j}(u,x,y,z)=e_{(j)i}(u,x,y,z)\;.
\label{20}
\end{equation}
For the flat space-time tetrad field Eq. (20) ensures that
the physical space-time with coordinates $x^\mu$ is not rotating
with respect to the reference space-time with coordinates $q^a$.
This condition is required to hold for a set of tetrad fields
that describe an arbitrary gravitational field \cite{Maluf1}.

The set of tetrad fields given by Eq. (17) is fixed precisely by
the six conditions determined by Eqs. (19) and (20). We propose that
the isotropic violation of the time gauge condition, given in the
present context by Eq. (19), together with the condition defined by
Eq. (20) fix the structure of any set of tetrad fields for  radiating
space-time metric tensors determined in terms of the retarded time
$u$, in which case the imposition of the time gauge condition is
troublesome. We argue that such tetrad field structure is adapted
to static observers at spacelike infinity (because of Eq. (20))
and is suitable to investigate the properties of the energy flux
of a system that emits gravitational waves, as we will see.

Before closing this section we note that the idea of a
{\it preferred frame} has been recently considered in connection
with investigations either in cosmology or in attempts to the quantum 
theory of gravity \cite{PF}. The motivation behind such
attempts to a preferred frame are quite different from our
purposes, and even the approach itself is conceptually and
mathematically different from the what has been presented above.
Nevertheless it is worth noting that similar ideas emerge in several
distinct analyses in general relativity.
\bigskip

\section{Bondi energy and the energy loss}

By considering $a=(0)$ in Eq. 7 we have an expression for the
energy of the gravitational field contained within a volume
$V$ of the three-dimensional spacelike hypersurface, bounded
by a surface $S$ of constant radius $r$,

\begin{eqnarray}
E&=&-\int_{V\rightarrow \infty} d^3x\,\partial_j
\Pi^{(0)j}\nonumber \\
&=&-\int_{V\rightarrow \infty} d^3x\, \partial_j
(-4ke\Sigma^{(0)0j})\nonumber \\
&=&-\int_{S\rightarrow \infty} dS_j(-4ke\Sigma^{(0)0j})\nonumber \\
&=&\int_{r \rightarrow \infty} d\theta d\phi(4ke \Sigma^{(0)01})\;,
\label{21}
\end{eqnarray}
The tensor $\Sigma^{(0)01} =e^{(0)}\,_0\Sigma^{001}+
e^{(0)}\,_j\Sigma^{j01}$ is evaluated by means of Eq. (2). It
simplifies to

\begin{eqnarray}
\Sigma^{(0)01}&=&{1\over 2}
e^{(0)}\,_0(T^{001}+g^{01}T^0)\nonumber \\
& & +{1\over 2} e^{(0)}\,_1 (T^{101}+g^{11}T^0-g^{01}T^1)\;.
\label{22}
\end{eqnarray}
We need to calculate all components $T_{\lambda \mu \nu}=
e^a\,_\lambda T_{a\mu\nu}$ of the torsion tensor obtained from
Eq. (17). For later purposes, we also provide the asymptotic
behavious of these componentes with respect to the radial
variable $r$, in the limit $r\rightarrow \infty$. The nonvanishing
components are given by

\begin{eqnarray}
T_{001}&=&{1\over 2}\partial_1(A^2-C^2)-A\,\partial_0 B
\;\simeq \; O({1\over r}) \nonumber \\
T_{101}&=&B\,\partial_1 A\; \simeq \; O({1\over r^2})\nonumber \\
T_{201}&=&-re^\gamma\partial_1 C\; \simeq \; O({1\over r})\nonumber \\
T_{002}&=&{1\over 2}\partial_2(A^2-C^2)+C(\partial_0 \gamma)
re^\gamma\; \simeq \; O({1\over r})\nonumber \\
T_{102}&=&B\,\partial_2 A+ BC\; \simeq \; O({1\over r})\nonumber \\
T_{202}&=&re^\gamma\lbrack (\partial_0 \gamma)re^\gamma
-\partial_2 C \rbrack\; \simeq \; O(r)\nonumber \\
T_{303}&=&-(\partial_0 \gamma)r^2e^{-2\gamma}\sin^2\theta-
re^{-\gamma}C\sin\theta\,\cos\theta\; \simeq \;O(r)\nonumber \\
T_{012}&=&A\,\partial_2 B+ C\lbrack e^\gamma+
(\partial_1 \gamma)re^\gamma -B \rbrack \;\simeq
\;O({1\over r})\nonumber \\
T_{212}&=&re^\gamma \lbrack e^\gamma + (\partial_1 \gamma)
re^\gamma -B \rbrack \; \simeq \; O(r)\nonumber \\
T_{313}&=&re^{-\gamma}\lbrack e^{-\gamma}-(\partial_1 \gamma)
re^{-\gamma} -B \rbrack \; \simeq \; O(r)\nonumber \\
T_{323}&=&-r^2(\partial_2 \gamma) e^{-2\gamma} \sin^2\theta-
r^2(1-e^{-2\gamma})\sin\theta\,\cos\theta\;
\simeq O(r^2)\;.\nonumber \\
\label{23}
\end{eqnarray}
We also need the traces $T^\mu$. They read

\begin{eqnarray}
T^0&=&g^{01}g^{01}T_{101}-g^{01}(g^{22}T_{212}+
g^{33}T_{313})\nonumber \\
T^1&=&-g^{01}g^{01}T_{001}+g^{01}g^{12}(T_{012}-T_{201})-
g^{11}g^{22}T_{212}-g^{11}g^{33}T_{313}\nonumber \\
& & -g^{01}g^{22}T_{202}-g^{01}g^{33}T_{303}
-g^{12}g^{33}T_{323}+g^{12}g^{12}T_{212}\nonumber \\
T^2&=&g^{01}g^{12}T_{101}-g^{12}g^{22}T_{212}-
g^{12}g^{33}T_{313}\nonumber \\
T^3&=&0
\label{24}
\end{eqnarray}

After the substitution of Eqs. (23) and (24) into Eq. (22) we obtain

\begin{eqnarray}
\Sigma^{(0)01}&=&-{1\over 2}e^{(0)}\,_0
\lbrack g^{01}g^{01}g^{22}T_{212}+
g^{01}g^{01}g^{33} T_{313}\rbrack\nonumber \\
& & +{1\over 2}e^{(0)}\,_1\lbrack g^{01}g^{01}g^{22}T_{202}+
g^{01}g^{01}g^{33}T_{303}+g^{01}g^{12}g^{33}T_{323}\rbrack \;.
\label{25}
\end{eqnarray}
With the help of the equation above we arrive at an expression for the
total gravitational energy of the space-time,

\begin{eqnarray}
E&=& {1\over {4\pi}}\lim_{r \rightarrow \infty}
\int^{2\pi}_0 d\phi \int^\pi_0 d\theta {1\over 2}
\biggl\{r \sin\theta \biggl[e^\gamma + e^{-\gamma} 
-2e^{-\beta}\biggl({V\over r}\biggr)^{1\over 2}\biggr]\nonumber \\
& &+r^2 e^{-\beta}\biggl( {V\over r}\biggr)^{-{1\over 2}}
{\partial \over {\partial \theta}}(U\sin\theta)\biggr\}\nonumber \\
&=& {1\over 4} \lim_{r \rightarrow \infty}\int^\pi_0 d\theta
\biggr\{ r\sin\theta\biggl[e^\gamma + e^{-\gamma}
-2e^{-\beta}\biggl({V\over r}\biggr)^{1\over 2}\biggr]\nonumber \\
& &+r^2e^{-\beta}\biggr( {V\over r}\biggr)^{-{1\over 2}}
{\partial \over {\partial \theta}}(U\sin\theta)\biggr\}\;.
\label{26}
\end{eqnarray}
Considering the last integral in Eq. (26), we note that

\begin{equation}
\lim_{r \rightarrow \infty} \int^\pi_0 d\theta\,
r^2 e^{-\beta}\biggl( {V\over r}\biggr)^{-{1\over 2}}
{\partial \over {\partial \theta}}(U\sin\theta)=
\lim_{r \rightarrow \infty} \int^\pi_0 d\theta\, r^2
{\partial \over {\partial \theta}}(U\sin\theta)\;.
\label{27}
\end{equation}
Since the function $U(\theta)\sin\theta$ is required to vanish
at $\theta= 0,\pi$ \cite{Bondi}, we conclude that the integral above
vanishes. Moreover we have

\begin{equation}
e^\gamma + e^{-\gamma}-2e^{-\beta}\biggl(
{V\over r} \biggr)^{1\over 2} \simeq
{{2M}\over r}- {c^2 \over r^2}\;,
\label{28}
\end{equation}
in the limit $r \rightarrow \infty$. Therefore in such limit we
obtain

\begin{equation}
E={1\over 2} \int^\pi_0 d\theta\, \sin\theta \,M(u,\theta)\;,
\label{29}
\end{equation}
which is precisely the expression of the Bondi energy.

It is possible to show by explicit calculations that the Bondi
energy is invariant under local Lorentz transformations,
$\tilde e^a\,_\mu(x)=\Lambda^a\,_b(x) e^b\,_\mu(x)$, whose asymptotic
behaviour is given by

\begin{eqnarray}
\Lambda^{(0)}\,_{(0)}&=& 1+\,\,^1\omega^{(0)}\,_{(0)}
(1 /r)\nonumber \\
\Lambda^{(0)}\,_{(i)}&=&\,\,^1\omega^{(0)}\,_{(i)}
(1 / r)\nonumber \\
\Lambda^{(i)}\,_{(j)}&=&\delta^{(i)}_{(j)}+\,\,
^0\omega^{(i)}\,_{(j)}+\,\,^1\omega^{(i)}\,_{(j)}(1/r)\;,
\label{30}
\end{eqnarray}
where $\,\,^0\omega_{(i)(j)}$ and $\,\,^1\omega_{ab}$ are
infinitesimal quantities such that
$\,\,^0\omega_{(i)(j)}  =-\,\,^0\omega_{(i)(j)}$, 
$\,\,^1\omega_{ab}  =-\,\,^1\omega_{ba}$, and
$\,\,^0\omega_{(i)(j)}$ are constants. This result is no surprise
since Eq. (7) can be evaluated as a surface integral, and all
transformations that preserve the asymptotic structure of the
field quantities are allowed. Nevertheless it is possible to show by
explicit calculations that in the limit $r \rightarrow \infty$ we
have

\begin{eqnarray}
\tilde T_{212}&=&T_{212} \nonumber \\
\tilde T_{313}&=&T_{313} \nonumber \\
\tilde T_{202}&=&T_{202} \nonumber \\
\tilde T_{303}&=&T_{303} \nonumber \\
\tilde T_{323}&=&T_{323} \;,
\label{31}
\end{eqnarray}
and therefore

\begin{eqnarray}
\tilde \Sigma^{(0)01}&=&
-{1\over 2}(\Lambda^{(0)}\,_{(0)} e^{(0)}\,_0+
\Lambda^{(0)}\,_{(i)} e^{(i)}\,_0 ) \lbrack
g^{01}g^{01}g^{22} T_{212}+g^{01}g^{01}g^{33} T_{313}
\rbrack \nonumber \\
& & +{1\over 2}(\Lambda^{(0)}\,_{(0)} e^{(0)}\,_1+
\Lambda^{(0)}\,_{(i)} e^{(i)}\,_1 )
\lbrack
g^{01}g^{01}g^{22} T_{202} + g^{01}g^{01}g^{33} T_{303}\nonumber \\
& &+g^{01}g^{12}g^{33} T_{323}\rbrack\;,
\label{32}
\end{eqnarray}
from what we conclude that Eq. (21) is invariant under the
transformations given by Eq. (30).
We note that in the expression above the tetrad components
$e^{(i)}\,_1$ behave as $r^0$ when $r \rightarrow \infty$.

The flux of gravitational energy is determined by the $a=(0)$
component of Eq. (9), and by making $j=1$, i.e., by integrating
over a surface $S$ of constant radius $r$, and requiring
$r \rightarrow \infty$. It reads

\begin{eqnarray}
\Phi^{(0)}_g&=&k\int_S dS_j\lbrack e e^{(0)\mu}
(4\Sigma^{bcj}T_{bc\mu}-\delta^j_\mu
\Sigma^{bcd}T_{bcd})\rbrack \nonumber \\
&=&k \int_S dS_1\, e\lbrack g^{01} e^{(0)}\,_0 \Omega^1\,_1 +
g^{01} e^{(0)}\,_1 \Omega^1\,_0 \nonumber \\
& & +g^{11} e^{(0)}\,_1 \Omega^1\,_1 +
g^{12} e^{(0)}\,_1 \Omega^1\,_2 \rbrack\;,
\label{33}
\end{eqnarray}
where

\begin{eqnarray}
\Omega^1\,_1&=& 4\Sigma^{\lambda \sigma 1}T_{\lambda \sigma 1}-
\Sigma^{\lambda \sigma \nu}T_{\lambda \sigma \nu}\;, \nonumber \\
\Omega^1\,_0&=&
4\Sigma^{\lambda \sigma 1}T_{\lambda \sigma 0}\;, \nonumber \\
\Omega^1\,_2&=& 4\Sigma^{\lambda \sigma 1}T_{\lambda \sigma 2}\;.
\label{34}
\end{eqnarray}

The expression of the quantities above are given in the Appendix.
In order to calculate the total flux in the limit
$r \rightarrow \infty$ we take into account the asymptotic
behaviour of the following field quantities,

\begin{eqnarray}
g^{01} & \rightarrow & -1\;, \nonumber \\
g^{11} & \rightarrow & 1\;, \nonumber \\
g^{12} & \rightarrow & O(1/r^2)\;, \nonumber \\
e^{(0)}\,_0 & \rightarrow & 1\;, \nonumber \\
e^{(0)}\,_0 & \rightarrow & 1 \;,\nonumber \\
e & \rightarrow & r^2\sin\theta\;.
\label{35}
\end{eqnarray}
It is easy to verify that in the limit $r \rightarrow \infty$ the
first and third term on the right hand side of Eq. (33) cancel
each other, and that the last term behaves as
$g^{12}e^{(0)}\,_1\Omega^1\,_2 =O(1/r^4)$. Therefore $\Phi^{(0)}_g$
reduces to

\begin{equation}
\Phi^{(0)}_g={1\over {16 \pi}}\int_{S \rightarrow \infty}
d\theta\, d\phi\, e(-\Omega^1\,_0)\;.
\label{36}
\end{equation}

In order to simplify the evaluation of the integral above we
only consider terms in Eq. (A.2) that fall off as $1/r^2$.
For this purpose we take into account Eqs. (23) and (35).
We conclude that five terms may possibly contribute to the surface
integral when $r \rightarrow \infty$. By integrating in $\phi$
we find that in this limit Eq. (36) simplifies to

\begin{eqnarray}
\Phi^{(0)}_g&=&{1\over 4}\int_{S\rightarrow \infty}d\theta\,
(r^2\sin\theta)\biggl[
g^{01}g^{01}g^{22}T_{001}T_{202}+g^{01}g^{01}g^{33}T_{001}T_{303}
\nonumber \\
& & g^{11}g^{22}g^{33}\biggl(T_{202}T_{313}+T_{303}T_{212}\biggr)
+2g^{01}g^{22}g^{33}T_{202}T_{303}\biggr]\;.
\label{37}
\end{eqnarray}
By explicitly calculating the asymptotic behaviour of all quantities
under integration, and after carrying out
several cancellations, we conclude that only one term arising from
the product $T_{202}T_{303}$ in Eq. (37) yields a nonvanishing
value to $\Phi^{(0)}$. We obtain

\begin{equation}
\Phi^{(0)}_g={1\over 2} \int_0^\pi d\theta\,
\sin\theta(\partial_0 c)^2\;,
\label{38}
\end{equation}
which is the well known value of the loss of mass in Bondi's
space-time \cite{Bondi}. We note that no multiplicative factors
had to be adjusted in Eqs. (9) or (33), in order to arrive at the
expression above. To our knowledge it is the first time that the
loss of mass has been obtained by means of a flux equation for the 
gravitational energy. Finally we note that since Eq.(14) is a
vacuum solution of Einstein's equations, then $\Phi^{(0)}_m=0$
and thus Eq. (11) is naturally verified.

\bigskip

\section{The Vaidya metric tensor}

The Vaidya metric \cite{Vaidya}
describes the exterior geometry of a spherically
symmetric star when the energy flux of matter is included. The star
is considered to be throwing away mass.
A spherically symmetric matter distribution does not emit
gravitational radiation by means of gravitational waves, as in
Bondi's space-time. Therefore in the present context the energy flux
is a purely matter flux. The metric is determined by the mass
function $m(u)$, which is a nonincreasing function of the retarded
time. It reads

\begin{equation}
ds^2=- ( 1-{{2m(u)}\over r} ) du^2-2du\,dr+
r^2(d\theta^2+\sin\theta^2\,d\phi^2)\;.
\label{39}
\end{equation}
If we require the function $m$ to be constant, then the metric
tensor above reduces to the Schwarzschild solution.

The set of tetrad fields that yields Eq. (39) is obtained
by following the same steps that led to Eq. (17). We require
$e^a\,_\mu$ to satisfy a condition similar to Eq. (19), i.e., the
isotropic violation of the time gauge condition, together with Eq.
(20). The tetrad field for which both conditions are satisfied is
given by

\begin{equation}
e_{a\mu}(u,r,\theta,\phi)=\pmatrix{-\alpha& -\alpha^{-1}&0&0\cr
0&\alpha^{-1}
\sin\theta\cos\phi & r\cos\theta \cos\phi&-r\sin\theta\sin\phi\cr
0&\alpha^{-1}
\sin\theta\sin\phi & r\cos\theta \sin\phi& r\sin\theta\cos\phi\cr
0&\alpha^{-1}
\cos\theta& -r\sin\theta& 0\cr}\;,
\label{40}
\end{equation}
where $\alpha=(1-2m/r)^{1/2}$. The nonvanishing components of
$T_{\mu\nu\lambda}=e^a\,_\mu T_{a\nu\lambda}$ are 

\begin{eqnarray}
T_{001}&=& -{{\dot m}\over {\alpha^2 r}}+{m \over r^2}\;,\nonumber \\
T_{101}&=& {m \over {\alpha^2 r^2}}\;,\nonumber \\
T_{212}&=& r(1-\alpha^{-1})\;, \nonumber \\
T_{313}&=& r(1-\alpha^{-1})\sin\theta^2\;,
\label{41}
\end{eqnarray}
with $\dot m =dm/du$. The calculations in this context are much
simpler than in the preceeding case. We have

\begin{eqnarray}
\Sigma^{(0)01}&=& e^{(0)}\,_0\Sigma^{001}+
e^{(0)}\,_1\Sigma^{101}\nonumber \\
&=& {1\over 2}e^{(0)}\,_0(T^{001}+g^{01}T^0) \nonumber \\
& &+{1\over 2}e^{(0)}\,_1(T^{101}+g^{11}T^0-g^{01}T^1)\;.
\label{42}
\end{eqnarray}
After a number of simplifications we obtain

\begin{eqnarray}
\Sigma^{(0)01}&=&
-{1\over 2} e^{(0)}\,_0(g^{22}T_{212}+g^{33}T_{313})\nonumber \\
&=& {1\over r}(1-\alpha)\;.
\label{43}
\end{eqnarray}

Therefore by considering a finite volume $V$, enclosed by a
spherical surface $S$ of radius $r$, we have

\begin{eqnarray}
E(u)&=&-\int_V d^3x\,\partial_j
\Pi^{(0)j}\nonumber \\
&=&\int_S d\theta d\phi(4ke \Sigma^{(0)01})\nonumber \\
&=&r\biggl( 1- \sqrt{ 1-{{2m(u)}\over r} } \biggr)\;.
\label{44}
\end{eqnarray}
The total gravitational energy of the space-time is obtained by
making $r \rightarrow \infty$. It is given by $E_{total}=m(u)$.
From Eq. (44) it follows that

\begin{equation}
{{dE(u)}\over {du}}= {{\dot m(u)}\over
\sqrt{1-{{2m(u)}\over r}}}\;.
\label{45}
\end{equation}

Now we proceed to evaluate the fluxes $\Phi^{(0)}_g$ and
$\Phi^{(0)}_m$. The former is given by

\begin{eqnarray}
\Phi^{(0)}_g&=&k\int_S dS_j\lbrack e e^{(0)\mu}
(4\Sigma^{bcj}T_{bc\mu}-\delta^j_\mu
\Sigma^{bcd}T_{bcd})\rbrack \nonumber \\
&=&k \int_S dS_1\, e\lbrack g^{01} e^{(0)}\,_0 \Omega^1\,_1 +
g^{01} e^{(0)}\,_1 \Omega^1\,_0  +
g^{11} e^{(0)}\,_1 \Omega^1\,_1 \rbrack\;,
\label{46}
\end{eqnarray}
where $S$ is a surface of constant radius $r$ and the quantities
$\Omega^\mu_\nu$ are defined by Eq. (34).
For the Vaidya metric tensor we have $g^{00}=0$, $g^{01}=-1$,
$g^{11}=\alpha^2$ and $g^{12}=0$. By taking into account the
nonvanishing components of $T_{\mu\nu\lambda}$ it is possible to
conclude that in the present case we find

\begin{eqnarray}
\Omega^1\,_0&=&0\;, \nonumber \\
\Omega^1\,_1&=& \Sigma^{\mu\nu\lambda}T_{\mu\nu\lambda}\;.
\label{47}
\end{eqnarray}
Moreover we have $g^{01} e^{(0)}\,_0=-\alpha$ and
$g^{11} e^{(0)}\,_1 =\alpha$, from what
we conclude that $\Phi^{(0)}_g=0$,
as expected.

In order to calculate the energy flux of matter  $\Phi^{(0)}_m$
we need the energy-momentum tensor associated to Eq. (39). It
reads \cite{Lind}

\begin{equation}
T_{\mu\nu}=-
{1\over {4 \pi r^2}}{{dm(u)}\over {du}}\delta^0_\mu \delta^0_\nu\;.
\label{48}
\end{equation}
For a volume $V$ enclosed by a surface of constant radius $S$ we
find

\begin{eqnarray}
\Phi^{(0)}_m&=&\int_V d^3x\,
\partial_j(ee^{(0)}\,_\mu T^{j\mu})\nonumber \\
&=&\int_S dS_1\;e e^{(0)}\,_\mu T^{1\mu} \nonumber \\
&=& \int_S d\theta\,d\phi\;(r^2\sin\theta)e^{(0)}\,_1 T^{11}\;.
\label{49}
\end{eqnarray}
Given that $T^{11}=-{\dot m}/(4\pi r^2)$, we obtain

\begin{equation}
\Phi^{(0)}_m=-{{\dot m} \over \sqrt{1-{{2m(u)}\over r}}}\;,
\label{50}
\end{equation}
Therefore Eqs. (45) and (50) are in agreement with Eq. (11). We
conclude that in the Vaidya space-time the energy flux of matter 
accounts for the loss of gravitational energy. By taking the
limit $r \rightarrow \infty$ in Eqs. (45) and (50) we recover the
well known result of the total flux (total luminosity $L_\infty$)
measured by an observer at infinity, $L_\infty = -dm/du$
\cite{Lind}.
\bigskip

\section{Concluding remarks}

The flux of the energy-momentum of gravitational and matter fields
has not been investigated so far in the literature in a unique
geometrical setting. In particular, to our knowledge the flux of
matter fields $\Phi^a_m$ defined by Eq. (10) has not been previously
considered. The total flux of matter fields does not depend on the
particular structure of the tetrad fields, but only on the asymptotic
behaviour of the latter. The remarkable feature of the continuity
equation (8) is that it is a natural consequence of
Einstein's equations written in the TEGR.
The detailed analysis of Eq. (11) in the context of Bondi and Vaidya
space-times demonstrates the consistency of the present
framework for investigating the energy-momentum properties of the
gravitational field. One may then consider the application of the
techniques presented here to concrete astrophysical problems, in
which case one may assume the surface $S$ in Eq. (9) to be an
{\it open} surface, as considered in Ref. \cite{Maluf3}. In the
latter reference we evaluated the energy flux of plane
gravitational waves. It is likely that for a given set of tetrad
fields a nonvanishing value of $\Phi^{(0)}_g$ indicates the emission
of gravitational waves. However, Bondi argues that the gravitational
energy can be transferred by means of two distinct ways, radiative
transfer and inductive transfer \cite{Bondi2}. Therefore it is
possible that a nonvanishing value of $\Phi^{(0)}_g$ also indicates
a form of inductive transfer of gravitational energy. This issue
will be further analyzed.

Restricting considerations to spherical surfaces $S$ of constant
radius, it is possible to show by explicit calculations that
for the Bondi space-time the
three momenta $P^{(i)}$ given by Eq. (7) vanish. However, we note
that $P^{(3)}$ vanishes only by placing the surface $S$
at spacelike infinity. For a spherical surface $S$ of finite radius
$r$, $P^{(3)}$ is nonvanishing. The physical meaning of a
nonvanishing $P^{(i)}$ is under investigation.

It is an interesting achievement of the present approach to the
energy-momentum of the gravitational field that the definition given
by Eq. (7) allows a unfied treatment of the Bondi and ADM
energies. For this purpose we have to fix suitable expressions for
the tetrad field in each case. Although the localization of the
gravitational energy depends on the choice of the tetrad field,
total quantities are in fact invariant under a wide class of local
Lorentz transformations.

\bigskip
\noindent {\bf Acknowledgements}\par
\noindent F. F. Faria is grateful to the Brazilian agency CAPES for
financial support.\par

\bigskip
\centerline{\bf APPENDIX}

\bigskip

For the Bondi space-time metric tensor the quantities
$\Omega^\mu\,_\nu$ are given by
         
$$\Omega^1\,_1=g^{01}g^{01}g^{22}\biggl({1\over 2}T_{012}T_{012}+
T_{012}T_{201}-{1\over 2}T_{201}T_{201}$$

$$-2T_{001}T_{212}-{1\over 2}T_{102}T_{102}\biggr)$$

$$-2g^{01}g^{01}g^{33} T_{001}T_{313}
+2g^{01}g^{12}g^{22} T_{012}T_{212}$$

$$+2g^{01}g^{12}g^{33}\biggl(T_{012}T_{313} -
T_{201}T_{313}\biggr)$$

$$+g^{12}g^{12}g^{22}T_{212}T_{212}+
2 g^{12}g^{12}g^{33}T_{212}T_{313}$$

$$-2g^{11}g^{22}g^{33}T_{212}T_{313}-
g^{22}g^{33}g^{33} T_{323}T_{323}\;,\eqno(A.1)$$

\bigskip

$$\Omega^1\,_0=g^{01}g^{01}g^{22}\biggl(
3T_{002}T_{102}+T_{002}T_{012}+T_{002}T_{201}$$

$$-2 T_{202}T_{001}\biggr) -2 g^{01}g^{01}g^{33}T_{001}T_{303}
+2g^{11}g^{12}g^{22}T_{102}T_{212}$$

$$+2g^{01}g^{12}g^{22}\biggl(T_{002}T_{212}+T_{202}T_{102}
+T_{202}T_{012}\biggr)$$

$$+2g^{01}g^{12}g^{33}\biggl( T_{002}T_{313}-T_{102}T_{303}
+T_{012}T_{303}$$

$$-T_{303}T_{201}\biggr) +2g^{01}g^{11}g^{22}\biggl(T_{102}T_{012}+
T_{102}T_{102}\biggr)$$

$$+2g^{12}g^{12}g^{33}\biggl(-T_{102}T_{323}+T_{202}T_{313}
+T_{303}T_{212}\biggr)$$

$$-2g^{11}g^{22}g^{33}\biggl( T_{202}T_{313}+T_{212}T_{303}\biggr)
-4g^{01}g^{22}g^{33}T_{202} T_{303}\;,\eqno(A.2)$$

\bigskip

$$\Omega^1\,_2=
g^{01}g^{01}g^{22}\biggl(-2T_{002}T_{212}-T_{201}T_{202}+
T_{202}T_{102}$$

$$+T_{202}T_{012}\biggr) +
2g^{01}g^{01}g^{33}\biggl(T_{102}T_{303} -T_{001}T_{323}
-T_{002}T_{313}\biggr)$$

$$+2g^{01}g^{12}g^{33}\biggl( -T_{202}T_{313}+T_{102}T_{323}
+T_{012}T_{323} -T_{201}T_{323}\biggr)   $$

$$-2g^{01}g^{22}g^{33}T_{202}T_{323}
-2g^{11}g^{22}g^{33}T_{212}T_{323}
+2g^{12}g^{12}g^{33}T_{212}T_{323}\;.\eqno(A.3)$$

\vskip 1.0cm



\begin{thebibliography}{99}

\bibitem{Landau}
L. D. Landau and E. M. Lifshitz, {\it The Classical Theory of
Fields} (Pergamon Press, Oxford, 1980).

\bibitem{Weinberg}
S. Weinberg, {\it Gravitation and Cosmology: Principles and
Applications of the General Theory of Relativity}
(Wiley, New York, 1972).

\bibitem{Goldberg}
J. N. Goldberg, ``Invariant Transformations, Conservation Laws and
Energy-Momentum", in {\it General Relativity and Gravitation: One
Hundred Years After the Birth of Albert Einstein}, edited by
A. Held, Vol. 1, p. 469 (Plenum, New York, 1980).

\bibitem{Nester1}
C.-M. Chen and J. M. Nester, Gravitation and Cosmology
{\bf 6}, 257 (2000); C. C. Chang, J. M. Nester and C.-M. Chen,
Phys. Rev. Lett. {\bf 83}, 1897 (1999).

\bibitem{ADM}
R. Arnowitt, S. Deser and C. W. Misner, in {\it Gravitation: an
Introduction to Current Research}, edited by L. Witten (Wiley,
New York, 1962).

\bibitem{Bondi}
H. Bondi, M. G. J. van der Burg, and A. W. K. Metzner, Proc.
R. Soc. London, Ser. A {\bf 269}, 21 (1962).

\bibitem{Maluf1}
J. W. Maluf, J. F. da Rocha-Neto, T. M. L. Tor\'{\i}bio and K. H.
Castello-Branco, Phys. Rev. D {\bf 65}, 124001 (2002).

\bibitem{Maluf2}
J. W. Maluf, J. Math. Phys. {\bf 36}, 4242 (1995).

\bibitem{Hehl}
F. W. Hehl, J. D. McCrea, E. W. Mielke and Y. Ne'eman, Phys.
Rep. {\bf 258}, 1 (1995).

\bibitem{Mol}
C. M\o ller, ``Tetrad Fields and Conservation Laws in
General Relativity", Proceedings of the International School
of Physics Enrico Fermi, edited by C. M\o ller (Academic
Press, London, 1962);
``Conservation Laws in the Tetrad Theory of
Gravitation", Proceedings of the Conference on Theory of
Gravitation, Warszawa and Jablonna 1962 (Gauthier-Villars,
Paris, and PWN-Polish Scientific Publishers, Warszawa, 1964)
(NORDITA Publications No. 136).

\bibitem{Maluf3}
J. W. Maluf, F. F. Faria and K. H. Castello-Branco, Class.
Quantum Grav. {\bf 20}, 4683 (2003).

\bibitem{Bondi2}
H. Bondi, Proc. R. Soc. Lond. A {\bf 427}, 249 (1990).

\bibitem{Synge}
J. L. Synge, {\it Relativity: the General Theory} (North Holland,
Amsterdam, 1960).

\bibitem{Aldrovandi}
R. Aldrovandi, P. B. Barros and J. G. Pereira, Found. Phys.
{\bf 33}, 545 (2003).

\bibitem{Vaidya}
P. C. Vaidya, Nature {\bf 171}, 260 (1953).

\bibitem{Lind}
R. W. Lindquist, R. A. Schwartz and C. W. Misner, Phys. Rev.
{\bf 137}, B1364 (1965).

\bibitem{Maluf4}
J. W. Maluf, J. Math. Phys. {\bf 35}. 335 (1994).

\bibitem{Maluf5}
J. W. Maluf and J. F. da Rocha-Neto, Phys. Rev. D {\bf 64},
084014 (2001).

\bibitem{BN}
M. Blagojevi\'c and I. A. Nikoli\'c, Phys. Rev. D {\bf 62}, 024021
(2000).


\bibitem{Maluf6}
J. W. Maluf and J. F. da Rocha-Neto, Gen. Rel. Grav. {\bf 31},
173 (1999).

\bibitem{Maluf7}
J. W. Maluf and J. F. da Rocha-Neto, J. Math. Phys. {\bf 40}, 1490
(1999).

\bibitem{PF}
A. Albrecht and J. Magueijo, Phys. Rev. D {\bf 59}, 043516 (1999);
J. Magueijo, Rept. Prog. Phys. {\bf 66}, 2025 (2003); M. Arminjon,
``Ether theory of gravitation: why and how?" [gr-qc/0401021];
M. Arminjon, ``Gravitational effects on light rays and binary pulsar
energy loss in a scalar theory of gravity", Theor. Math. Phys.
(Teor. Mat. Fiz.), to appear [gr-qc/0301062]; M. Arminjon,
``Point-particle limit in a scalar theory of gravitation and the
weak equivalence principle" [gr-qc/0306025]; M. Consoli and
E. Costanzo, ``The motion of the solar system and the
Michelson-Morley experiment"  [astro-ph/0311576].


\end{thebibliography}
\end{document}